\documentclass[aps,prb,amsmath, amssymb, english, twocolumn,reprint,floatfix, superscriptaddress]{revtex4-1}

\usepackage[english]{babel}
\usepackage{graphicx}
\usepackage{verbatim}

\usepackage{epsfig}
\usepackage{epstopdf}
\usepackage{amsfonts}
\usepackage{amsthm}
\usepackage{amsmath}
\usepackage{amssymb}
\usepackage{color}
\usepackage[usenames,dvipsnames,svgnames,table]{xcolor}
\usepackage{hyperref} 
\hypersetup{backref,pdfpagemode=FullScreen,colorlinks=true,linkcolor=NavyBlue,citecolor=BrickRed}
\usepackage{makeidx}
\usepackage{pifont}
\usepackage{pdfpages}

\usepackage{color}
\usepackage{grffile}


\def\expect#1{\mathinner{\langle{#1}\rangle}}

{\catcode`\|=\active
  \gdef\expect#1{\left<\mathcode`\|"8000\let|\bravert {#1}\right>}}
\def\bravert{\egroup\,\vrule\,\bgroup}

\def\beq{\begin{equation}}
\def\eeq{\end{equation}}
\def\be{\begin{equation}}
\def\ee{\end{equation}}

\def\iomn{i\omega_n}

\def\cG0{{\cal G}_0}

\def\spinup{\uparrow}
\def\spindown{\downarrow}

\def\eps{\epsilon}

\def\s{\sigma}

\def\uc2{$U_{c2}$}
\def\uc1{$U_{c1}$}

\def\bea{\begin{eqnarray}}
\def\eea{\end{eqnarray}}
\def \bal{\begin{align}}
\def \eal{\end{align}} 
\def\#{\!\!}
\def\@{\!\!\!\!}

\def\+{\dagger}

\def\up{\spinup}
\def\down{\spindown}

\begin{document}

\title{\bf Chromium analogues of Iron-based superconductors}
\author{Martin Edelmann}
\author{Giorgio Sangiovanni}
\affiliation{Institut f\"ur Theoretische Physik und Astrophysik, Universit\"at W\"urzburg, Am Hubland, D-97074 W\"urzburg, Germany}
\author{Massimo Capone}
\affiliation{International School for Advanced Studies (SISSA), Via Bonomea 265, I-34136 Trieste, Italy}
\affiliation{CNR-IOM-Democritos National Simulation Centre, UOS Trieste-SISSA, Via Bonomea 265, I-34136 Trieste, Italy}
\author{Luca de'~Medici}
\affiliation{European Synchrotron Radiation Facility, 71 Av. des Martyrs, Grenoble, France}
\affiliation{Laboratoire de Physique et Etude des Mat\'eriaux, UMR8213 CNRS/ESPCI/UPMC, Paris, France}

\begin{abstract}
We theoretically investigate the $d^4$ (Cr$^{2+}$) compound BaCr$_2$As$_2$ and show that, despite non-negligible differences in the electronic structure, its many-body physics mirrors 
that of BaFe$_2$As$_2$, which has instead a $d^6$ (Fe$^{2+}$) configuration. This reflects a symmetry of the electron correlation effects around the half-filled $d^5$ Mott insulating state.
The experimentally known metallic antiferromagnetic phase is correctly modeled by dynamical mean-field theory, and for realistic values of the interaction it shows a moderate mass enhancement of order $\sim$2. This value decreases if the ordered moment grows as a result of a stronger interaction.
The antiferromagnetic phase diagram for this $d^4$ shows similarities with that calculated for the $d^6$ systems. 
Correspondingly, in the paramagnetic phase the influence of the half-filled Mott insulator shows up as a crossover from a weakly correlated to an orbitally differentiated ``Hund's metal'' phase which reflects an analogous phenomenon in $d^6$ iron compounds including a strong enhancement of the compressibility in a zone just inside the frontier between the normal and the Hund's metal. 
The experimental evidence and our theoretical description place BaCr$_2$As$_2$ at interaction strength slightly below the crossover which implies that negative pressures and/or electron doping (e.g. Cr $\rightarrow$ Mn,Fe or Ba $\rightarrow$ Sc,Y,La) might strongly enhance the compressibility, thereby possibly inducing a pairing instability in this non-superconducting compound.
\end{abstract}

\maketitle
\section{Introduction}
Since their discovery in 2008\cite{Kamihara_pnictides2}, Iron-based superconductors (FeSC) have polarized the attention in the field of high-temperature superconductivity. 
The many families that have been synthesized since then share several properties, starting from parent compounds showing a structural (nematic) transition between a high-temperature tetragonal and a low-temperature orthorhombic phase, in which often a striped-antiferromagnetic metallic phase is realized. Typically doping or pressure bring the system out of this low-T phase and superconductivity emerges.

Density-functional theory (DFT)  is reasonably successful in modeling the Fermi-surface topology of the parent compounds, most of which are semi-compensated metals showing hole pockets around the Brillouin zone center and electron pockets at the borders. This success, and the experimental evidence of Fermi-liquid behavior\cite{Rullier_Review}, supports a widespread view of these superconductors as weakly-correlated metals where superconductivity emerges thanks to a weak coupling 
to a low-energy bosonic mode. Electron-phonon coupling was quickly excluded from the candidates\cite{Boeri_e-ph_LaFeAsO} and to date spin fluctuations\cite{Mazin_Splusminus,Chubukov_ItinerantScenario,Fernandes_DriverSeat,Chubukov_Hirschfeld-PhysToday} or orbital fluctuations \cite{Kontani} are the most considered pairing mediators. The former is naturally suggested by the rather common proximity of superconductivity to a magnetically ordered phase, the latter is tied to a more elusive orbital ordering, which is a possible explanation of the phenomenology of those materials (like FeSe) where the magnetically ordered phase is not found.

However the rather bad metallicity\cite{Rullier_Review} of these compounds, together with some unsatisfactory aspect of the DFT modeling, such as overestimates of the band dispersion by a factor 2-3 (with corresponding discrepancies with the measured specific heat Sommerfeld coefficients, ARPES and quantum oscillation bandstructures) and of the ordered magnetic moments, naturally leads to include the electron correlation which is neglected in DFT and to a large extent in the perturbative approaches which attribute the superconducting and magnetic symmetry breaking to Fermi-surface instabilities.

A rather extreme point of view, where the low-energy physics is described in terms of a  $t$-$J_1$-$J_2$ model, was indeed considered since the very beginning\cite{Si_Abrahams_J1J2}, but faces the difficulty due to the absence of a Mott insulating state in the immediate proximity of the superconducting phase, which is a necessary starting point in this scenario. An intermediate perspective of coexisting weakly correlated and strongly correlated electrons was suggested in Ref. \onlinecite{demedici_3bandOSMT} based on orbital-selective Mott physics and phenomenologically assumed in models for magnetism and pairing\cite{Hackl_Vojta_OSMT_pnictides,Kou_OSMT_pnictides,Yin_Weiguo-Spin_fermion}.

In order to account in an unbiased way for the competition between the kinetic energy and the local interactions we include  multi-orbital Hubbard terms in the Hamiltonian.
These can be treated with a plethora of methods, from the exact solution of small cluster of atomic sites, to dynamical mean-field theory (DMFT) in the thermodynamic limit. Early studies showed that for realistic values of the interactions, sizable correlation effects should be present in this class of materials. In particular the Hund's coupling, the measure of the difference in electronic repulsion that electrons feel depending on their orbital character and spin alignment (giving rise to the well known ``Hund's rules'' in isolated atoms) was shown\cite{Haule_pnictides_NJP} to have a key effect. This is a particular manifestation of a very general phenomenology\cite{demedici_MottHund,demedici_Janus}, particularly strong in materials where the conduction bands arise from near-half-filled atomic shells, as it happens in FeSC where 5 conduction bands of mainly Iron character host 6 electrons per site in the parent compounds.
This phenomenology is reviewed in Ref. \cite{Georges_annrev}, while the metallic phase dominated by the Hund's coupling was called a ``Hund's metal''\cite{Yin_kinetic_frustration_allFeSC}.

Model studies identified a well-defined crossover line separating a normal metal from such Hund's metal in a density-interaction phase diagram and have characterized the latter 
in terms of high-spin local configurations and reduced inter-orbital correlation between charge fluctuations (For a review see Ref. \onlinecite{demedici-SpringerBook}).This phase can be roughly identified at finite temperatures with the so-called ``spin-freezing'' regime\cite{Werner_spinfreezing}.
This frontier of the Hund's metal phase departs from the Mott transition point at half-filling  and extends to finite doping bending towards larger interactions\footnote{
We recall that because of the Hund's coupling the Mott transition at half-filling is pushed in the range of the realistic interactions for FeSC, while at $n=5 \pm 1$ it is pushed to very large values of $U$, as it is illustrated, e.g., in Fig. of Ref. \onlinecite{demedici-SpringerBook} together with the above mentioned crossover line}.
The picture is confirmed also by realistic calculations  for FeSC\cite{Ishida_Mott_d5_nFL_Fe-SC,demedici_OSM_FeSC,demedici-SpringerBook,Liebsch_FeSe_spinfreezing,Werner_dynU_122,Misawa_d5-proximity_magnetic},
highlighting how the proximity of the  $d^5$ Mott insulating state determines  the strong correlations of the Hund's metal phase all the way up to the  $d^6$ FeSC parent compounds.

Since the many-body physics involved in these studies is symmetric around the half-filling $d^5$ atomic configuration, it is natural to expect that it should be realized also on the other side of the Mott insulating state, i.e. for $d^4$ compounds. In materials where the bandstructure is not too different from that of FeSC one can thus hope to find some of the phenomenology of the FeSC, among which high-temperature superconductivity.

An ideal candidate to verify these expectations is  BaCr$_2$As$_2$, which is analogous to BaFe$_2$As$_2$ in all aspects but the transition metal valence, that goes from $d^6$ for iron to $d^4$ for chromium. In this manuscript we perform a theoretical investigation of this material, explicitly including many-body effects.

The paper is organized as follows: In Section \ref{AF-DMFT} the DMFT results in the magnetically ordered phase are presented. Section \ref{paraSlSp} is instead dedicated to the paramagnetic phase investigated with the slave-spin mean-field method. In the last section we conclude with some considerations based on the result of our study.

\section{DMFT study of the G-type magnetic metal} \label{AF-DMFT}

BaCr$_2$As$_2$ was experimentally characterized in Ref. \onlinecite{Singh_BaCr2As2}, where resistivity and specific heat measures on single crystals were reported. In the same work density functional theory (DFT) calculations were performed to infer its magnetic state, that was identified as a G-type metallic antiferromagnet. This metal shows non-negligible correlation effects as indicated by the measured low-temperature slope of the specific heat, i.e. the Sommerfeld coefficient $\gamma$, which is a factor of two larger than the one calculated in DFT, as reported in Table \ref{TableSomm}. 
In the same table we report $\gamma$ calculated also with GGA and GGA+U. While the former yields, as expected, essentially the same result as LDA, the latter goes -- at a first sight surpisingly -- in the wrong direction compared to experiments.  
\begin{table}[h]
\begin{center}
\label{TableSomm}
\begin{tabular}{|c|c|}
\hline
 &  $\gamma$ [ mJ/(K$^2$ mol)] \\
\hline
experiment (from Ref. \onlinecite{Singh_BaCr2As2}) &    19.3 \\
\hline
LDA (from Ref. \onlinecite{Singh_BaCr2As2})  &     9.3  \\
\hline
GGA  (this work) &     9.4  \\
\hline
GGA+U (this work)  &     7.2  \\
\hline
DFT+DMFT (this work)  &     14.5 $\pm$ 0.5 \\
\hline
\end{tabular}
\end{center}
\caption{Values of the Sommerfeld coefficient $\gamma$ in BaCr$_2$As$_2$. The error bar in the last line comes from the fitting procedure required to esitimate the Fermi-liquid parameters from DMFT. The DMFT result refers to the density-density Hamiltonian.}
\end{table}

With GGA+U we indeed obtain an even smaller value of $\gamma$, as reported in Table \ref{TableSomm}.
The rationale for this is that while shifting the majority and minority densities of states, DFT+U does not describe the dynamical electronic correlations, which are instead expected to yield a mass enhancement and quasiparticle-bandstructure renormalization. This in turn leads to an enhancement of the total density of the states $D(\eps)$ and thus of the Sommerfeld coefficient $\gamma={\pi^2 k_B^2}/{3} \, D(\eps_F)$.
We thus study this compound with DFT+DMFT\cite{Kotliar_LDA_rmp} in order to incorporate this effect and properly explore the evolution of this magnetic phase with the interaction strength. 

\begin{figure}[h!]
\begin{center} 
  \includegraphics[width=7cm]{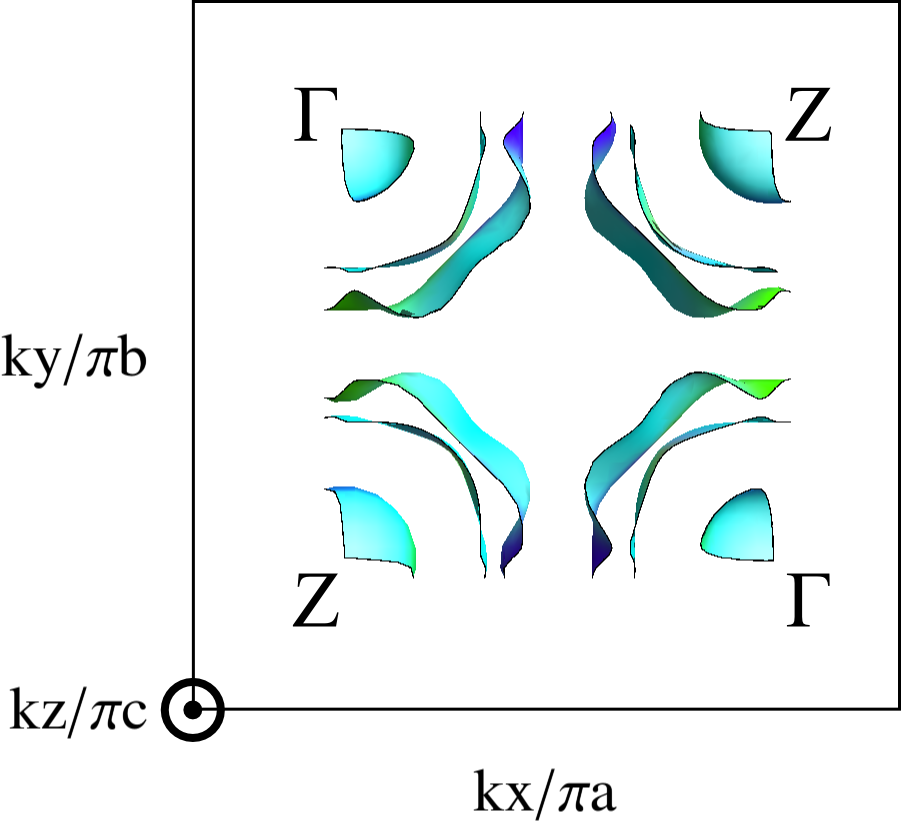}
    \includegraphics[width=7cm]{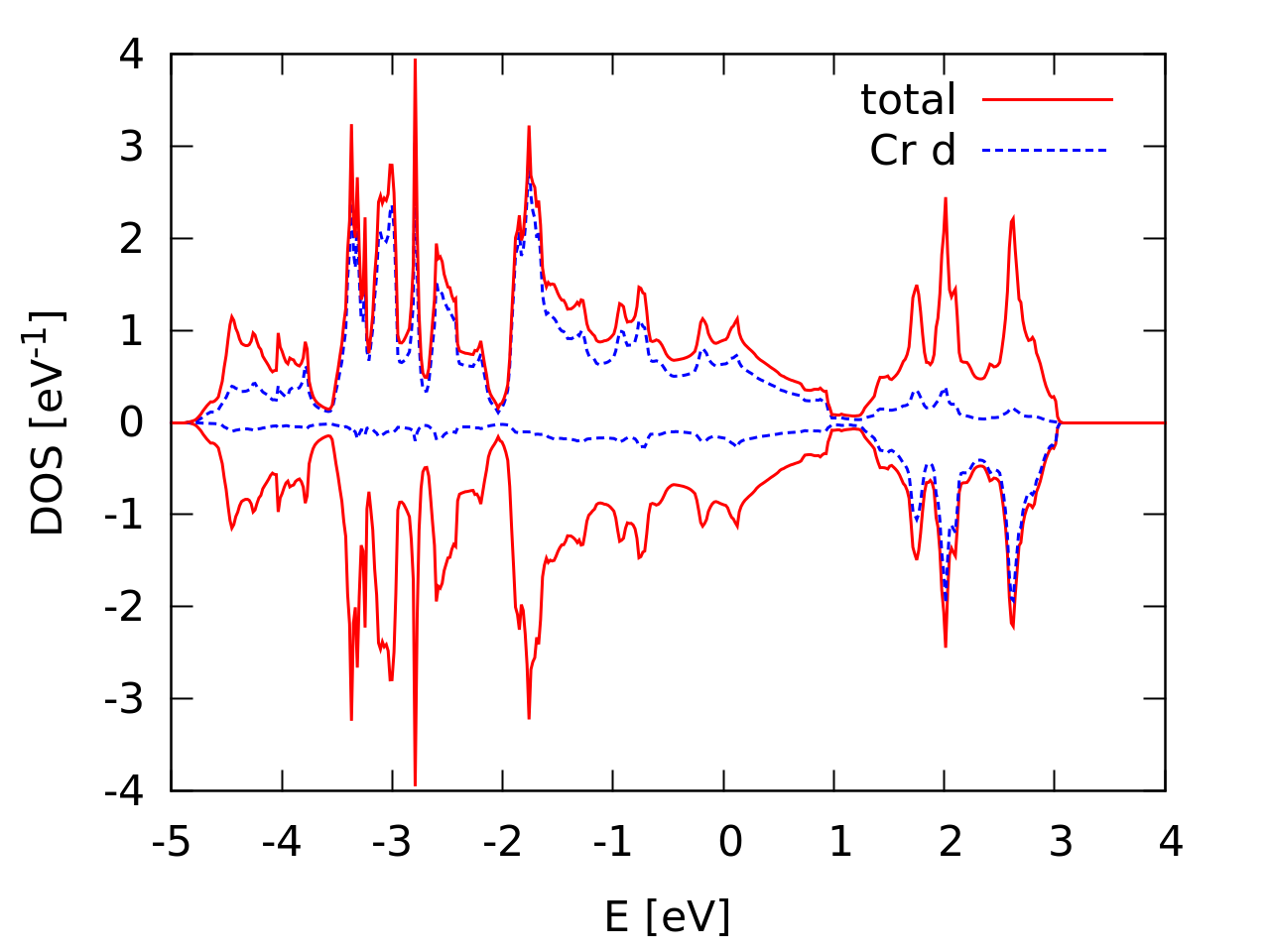}  
      \caption{GGA+U Fermi surface and DOS in the antiferromagnetic phase. The Fermi surfaces are colored according to the Wannier orbital contribution to the band creating said surface. Green shading indicates contrubition from a $xz$-like orbital while blue indicates contribution from a $yz$-like orbital. Teal shading thus signals a mixture of the two aforementioned contributions. The DOS is given in an energy range around the Fermi energy $\epsilon_\text{F}=0$. The partial DOS stemming from Cr-$d$ orbitals is indicated by the dashed blue line, with majority spin partial DOS on the positive $y$-axis and minority spin on the negative $y$-axis, respectively.}
  \label{fig:GGA+U}
\end{center}
\end{figure}

Before doing so, let us give some details about the DFT calculations:
The GGA results have been obtained using the projector augmented wave (PAW) method as implemented in the VASP package \cite{VASP1,VASP2}. For the lattice structure of BaCr$_2$As$_2$ we relied on data given Ref. \onlinecite{Singh_BaCr2As2} for the G-AFM phase. A 12$\times$12$\times$12 Monkhorst-Pack reciprocal lattice mesh as well as a plane wave energy cut-off of 283.9 eV has been used.
Electronic correlation in GGA+U calculations has been introduced via the rotationally invariant form introduced by Liechtenstein {\it et al.} \cite{LDAU} where the interaction strength for the Cr $d$-orbitals has been set to $U=F^0=2.112$ eV and $J=1/14(F^2+F^4)=0.602$. These values correspond to the cRPA results for BaFe2As2 ($U=2.8$eV, $J=0.43$eV, from Ref. \onlinecite{miyake_interactions_jpsj_2010}).
The obtained lowest energy phase (see table \ref{tab:GGA+U}) is as in the DFT case a G-type antiferromagnetic metal, of which we report the Fermi surfaces and the density of the states (DOS)
in Fig. \ref{fig:GGA+U}. The shape of these surfaces only differs marginally from those reported for LSDA calculations by Singh, {\it et al.} in Ref. \onlinecite{Singh_BaCr2As2}. 
\begin{table}[h]
\begin{center}
\begin{tabular}{|c|c|}
\hline
order & $E_0$ [$10^2$ eV] \\
\hline
PM: &       -0.28514224  \\
\hline
FM:   &    -0.29885715 \\
\hline
G-AFM: & -0.30235100 \\
\hline
\end{tabular}
\end{center}
\caption{Respective ground-state energies $E_0$ for the paramagnetic (PM), ferromagnetic (FM) and G-type antiferromagnetic (G-AFM) phases of BaCr$_2$As$_2$, calculated within DFT(GGA)+U, showing that the stable phase is the G-AFM as also obtained within DFT in Ref. \onlinecite{Singh_BaCr2As2}.}
\label{tab:GGA+U}
\end{table}

\begin{figure}[h!]
\begin{center} 
  \includegraphics[width=8.5cm]{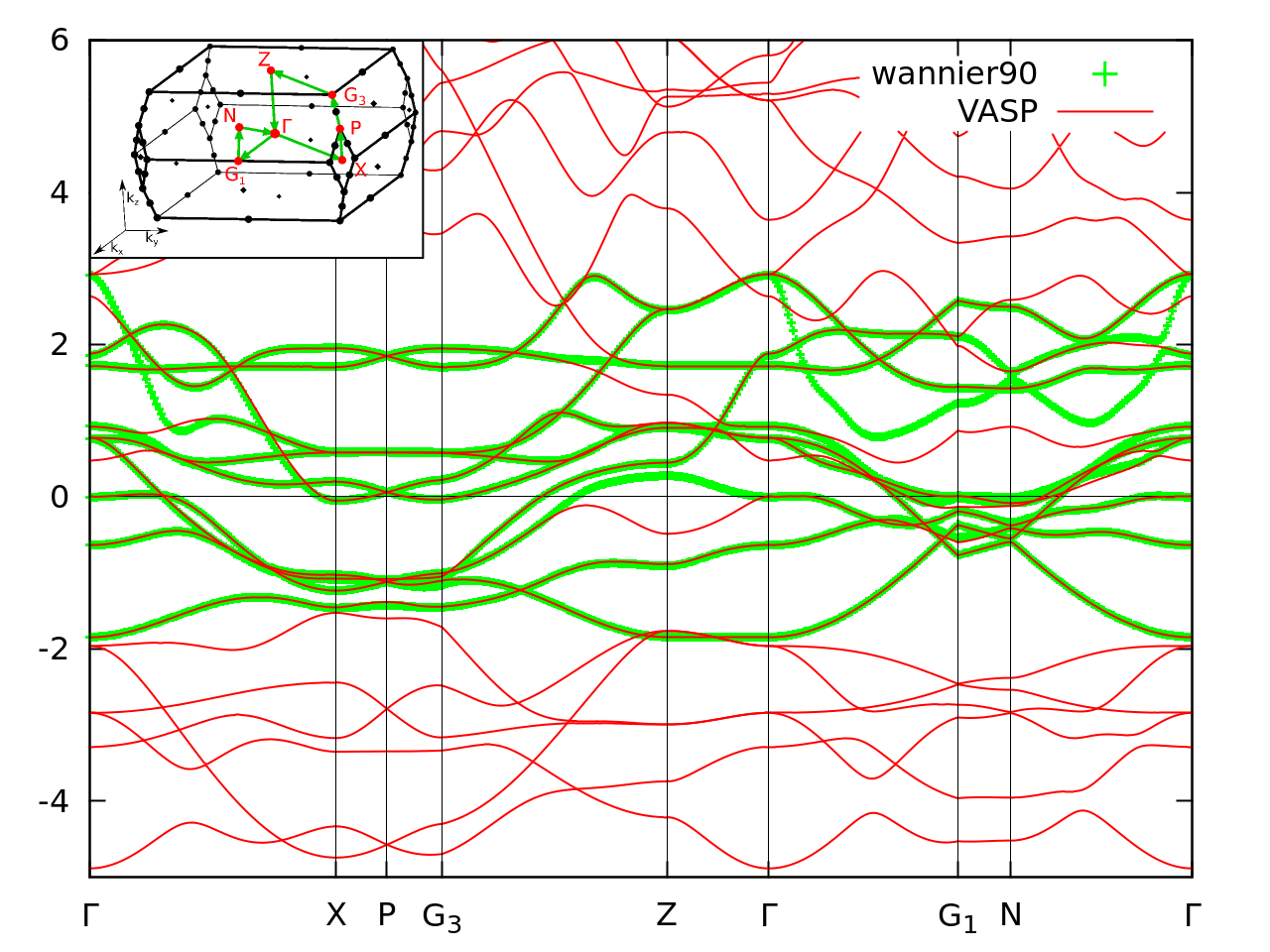} 
  \caption{Conduction bands (green lines) for BaCr$_2$As$_2$ calculated in GGA in the paramagnetic phase. The eigenvalues of the fitted Hamiltonian in the basis of Cr-centered $d$-like Wannier orbitals are shown in red.}
  \label{fig:Wannier}
\end{center}
\end{figure}

The dynamical treatment of the local electronic interaction is obtained supplementing an explicit (multi-orbital Hubbard-like) interaction term to the non-interacting Hamiltonian obtained by fitting the conduction bands of mainly Fe 3$d$-character. This fit (reported in Fig. \ref{fig:Wannier}) is done by expanding the relevant Bloch wave functions on an optimized one-particle basis of maximally localized Wannier orbitals (through the Wannier90 code \cite{Mostofi_Wannier90}).
The minimal basis is obtained by projection onto ten local Wannier orbitals, including the five 3$d$-orbitals on each of the two Cr atoms in the unit cell. Careful disentanglement of the Cr $d$-bands from those bands of mainly Ba $s$-character above and those of mainly As-$p$ character below in energy is necessary to find optimal agreement between the bands of the minimal local basis and the VASP bands. The strong hybridization between the Cr-$d$ and Ba-$s$ as well as As-$p$ bands in certain parts of the Brillouin zone leads to obvious deviations of the minimal basis eigenenergies to the DFT results in these regions. Fig. \ref{fig:Wannier} displays the best agreement between those two. The resulting local orbitals are used to construct the many-body basis states on which act the local interaction terms of a standard Kanamori Hamiltonian:
 \bea\label{H_int}
 H_{int}\#&\,=\,\#&U \sum_{il}
 n_{il\up} n_{il\down}+(U-2J)\sum_{i,l>l^\prime,\s} n_{il\s} n_{il^\prime\bar\s} \nonumber \\ 
  \#&\,+\,\#& (U-3J)\sum_{i,l>l^\prime,\s} n_{il\s}  n_{il^\prime\s}\nonumber \\
 \#&-\#&J\sum_{i,l\neq l^\prime}\left[d^\+_{il\up}d_{il\down}d^\+_{il^\prime\down}d_{il^\prime\up}+d^\+_{il\up}d^\+_{il\down}d_{il^\prime\up}d_{il^\prime\down}\right].
 \eea

A correction is in principle necessary to avoid double counting the static part of this interaction already included in the DFT Hamiltonian. However since we use here a basis of 5 Wannier d-like orbitals centered on each Cr atom this correction can simply be reabsorbed in an effective chemical potential which is fixed by imposing the correct total population of the conduction bands i.e. four electrons per Cr, in the stoichiometric compound. 
DMFT maps this lattice problem on a multi-orbital impurity model supplemented by a self-consistency condition\cite{georges_RMP_dmft}. 
We solve the impurity model here by continuous-time quantum Monte Carlo\cite{gullRMP,w2dynamics}. Most of the times we drop, unless explicitly state otherwise, the last two terms in Eq. (\ref{H_int}) and retain only the density-density terms of the interaction. This leads to a numerically much more tractable problem allowing us to explore the parameter space.

The spin- and orbitally-resolved self-energies $\Sigma_{m\s}(\iomn)$ (where  $\s=\uparrow,\downarrow$, $m$ indicates one of the five Wannier orbitals and $\iomn$ the fermionic Matsubara frequencies) have been calculated for both forms of the Hamiltonian, i.e. with the last two terms (Kanamori form) and without (density-density form) with $U=2.8$eV and $J=0.43$eV.
We have done so also without allowing for the antiferromagnetic long-range order, i.e. in the paramagnetic phase. 
When the imaginary part of these Matsubara self-energies are of the Fermi-liquid form (i.e. when they have a small imaginary part at zero frequency and a linear behavior) we have extracted the corresponding Fermi-liquid parameters, such as the renormalization of the quasiparticle bands $Z_m=(1-\partial\text{Im}\Sigma_m(\iomn)/\partial\iomn |_{\iomn \! \rightarrow \! 0})^{-1}$ as well as the extrapolated value at zero frequency which is proportional to the scattering rate of the quasiparticles. 
The calculated values are reported in Table \ref{tab:Z_G-Type} and in Table \ref{tab:Z_para} for the antiferromagnetic and paramagnetic phase, respectively.
\begin{table}[h]
\begin{center}
\begin{tabular}{|c|c|c|c|c|c|c|c|c|}
\multicolumn{1}{c}{}  & \multicolumn{4}{c}{$Z_m$ (spin $\uparrow$)} & \multicolumn{4}{c}{$Z_m$ (spin $\downarrow$)}  \\
\hline
orbital  & $z^2$ & $xz/yz$ & $xy$ & $x^2\#\!-\#y^2$ & $z^2$ & $xz/yz$ & $xy$ & $x^2\#\!-\#y^2$\\
\hline
Kanamori &    0.55 & 0.47 & 0.36 & 0.59 & 0.57 & 0.68 & 0.68 & 0.68\\
\hline
dens-dens   &    0.77  & 0.75  & 0.70  & 0.79  & 0.83 & 0.84  & 0.83  & 0.86 \\
\hline
\multicolumn{9}{c}{}  \\
\multicolumn{1}{c}{}  & \multicolumn{4}{c}{Im$\Sigma_m(i\omega_n \! \rightarrow \! 0)$ (spin $\uparrow$)} & \multicolumn{4}{c}{Im$\Sigma_m(i\omega_n \! \rightarrow \! 0)$ (spin $\downarrow$)}  \\
\hline
orbital & $z^2$ & $xz/yz$ & $xy$ & $x^2\#\!-\#y^2$ & $z^2$ & $xz/yz$ & $xy$ & $x^2\#\!-\#y^2$\\
\hline
Kanamori &    0.09 & 0.05 & 0.08 & 0.03 & 0.01 & 0.01 & 0.01 & 0.03\\
\hline
dens-dens   &    0.00  & 0.00  & 0.00  & 0.00  & 0.00 & 0.00  & 0.00  & 0.00 \\
\hline
\end{tabular}
\end{center}
\caption{Orbital- and spin- resolved quasiparticle weight (inverse mass enhancement) and values in eV of Im$\Sigma_m(i\omega_n \! \rightarrow \! 0)$ calculated in DFT+DMFT at $\beta = 100 eV^{-1}$ for the G-type antiferromagnetic phase. These numbers are extracted by means of a polynomial fit of the self-energies, excluding the first one Matsubara point from the fit in the cases where these display small deviations from the linear Fermi-liquid behaviour at the lowest frequencies.} 
\label{tab:Z_G-Type}
\end{table}

\begin{table}[h]
\begin{center}
\begin{tabular}{|c|c|c|c|c|}
\multicolumn{1}{c}{}  & \multicolumn{4}{c}{$Z_m$} \\
\hline
orbital & $z^2$ & $xz/yz$ & $xy$ & $x^2\#\!-\#y^2$ \\
\hline
Kanamori  &   0.39  &  0.43  &  0.47   &  0.47   \\ 
\hline
dens-dens   &  -  &  0.61  &  0.55  &  0.73   \\
\hline
\multicolumn{5}{c}{}  \\
\multicolumn{1}{c}{}  & \multicolumn{4}{c}{Im$\Sigma_m(i\omega_n \! \rightarrow \! 0)$ } \\
\hline
orbital & $z^2$ & $xz/yz$ & $xy$ & $x^2\#\!-\#y^2$ \\
\hline
Kanamori  &   0.55  &  0.16  &  0.37   &  0.11 \\
\hline
dens-dens   &  -  &  0.52  &  0.68  &  0.44 \\
\hline
\end{tabular}
\end{center}
\caption{Orbital-resolved quasiparticle weight (inverse mass enhancement) and values in eV of Im$\Sigma_m(i\omega_n \! \rightarrow \! 0)$ in the paramagnetic phase, calculated at $\beta = 100 eV^{-1}$ with DFT+DMFT. As in the antiferromagnetic case, these results have been obtained via a polynomial fit of the self-energies. No value means that, at this temperature, the DMFT Im$\Sigma_m(i\omega_n)$ displays too strong deviations from a linear and decreasing behavior at small frequencies that we did not attempt to extract the corresponding Fermi-liquid parameters.}
\label{tab:Z_para}
\end{table}

The Kanamori Hamiltonian yields thus a factor-of-two renormalization of the quasiparticle bands in the antiferromagnetic metallic phase, as indeed found in experiments\cite{Singh_BaCr2As2}. The corresponding Sommerfeld coefficient we obtain ($\sim$ 30 mJ/(K$^2$ mol)) is however higher than the measured one, but this is to some extent due to a small inaccuracy of our Wannier fit, which is unable to capture all the details of the low-energy band structure and has an extra pocket on the Fermi surface (as visible in Fig. \ref{fig:Wannier}) leading to a larger $D(\eps_F)$ compared to the fitted DFT one. 
This is not a severe drawback however, in the present context in which we are not investigating the detailed features of the Fermi surface. We are more interested in the many-body (Mott) physics due to local correlations, which is rather insensitive to these aspects of the band structure. We thus conclude that the interaction values that we used here, calculated for BaFa$_2$As$_2$, are roughly appropriate (or just slightly larger) for the description of BaCr$_2$As$_2$.
This is consistent with the fact that the two compounds are extremely similar in the structural parameters and the only difference is the transition metal ion. We thus find very similar radial extension of Wannier functions, just slightly superior for the Cr compound, consistent with its smaller atomic charge. Assuming reasonably a likewise similar screening of the electronic interaction, this points to interaction parameters in BaCr$_2$As$_2$ similar to BaFe$_2$As$_2$ or marginally smaller.

Remarkably, as it is clear from these tables, the density-density Hamiltonian yields somewhat less correlated results compared to the Kanamori one, for the same interaction parameters. This may come as a surprise as the experience in the paramagnetic phases is rather the opposite (the lower degeneracy of the atomic ground states in the density-density case favoring stronger correlations than in the Kanamori case).
This outcome can however be easily understood in light of the results that will be illustrated hereafter. Indeed we show how a stronger magnetization lowers the amount of mass renormalization. This in turn implies that, since the simplified form of the density-density Hund's coupling reduces the local spin fluctuations, a higher magnetization is thus favored and a weaker imaginary part of $\Sigma_m(i\omega_n)$ than with Kanamori is to be expected in the magnetically ordered phases. 
From Table \ref{tab:Z_para} one sees that the situation is reverted in the paramagnetic phase, where the density-density approximation gives much smaller values of $Z_m$ and in some cases even non-Fermi liquid results (for which we did not extract the value of $Z_m$ at all). 

\begin{figure}[h!]
\begin{center} 
  \includegraphics[width=4.25cm]{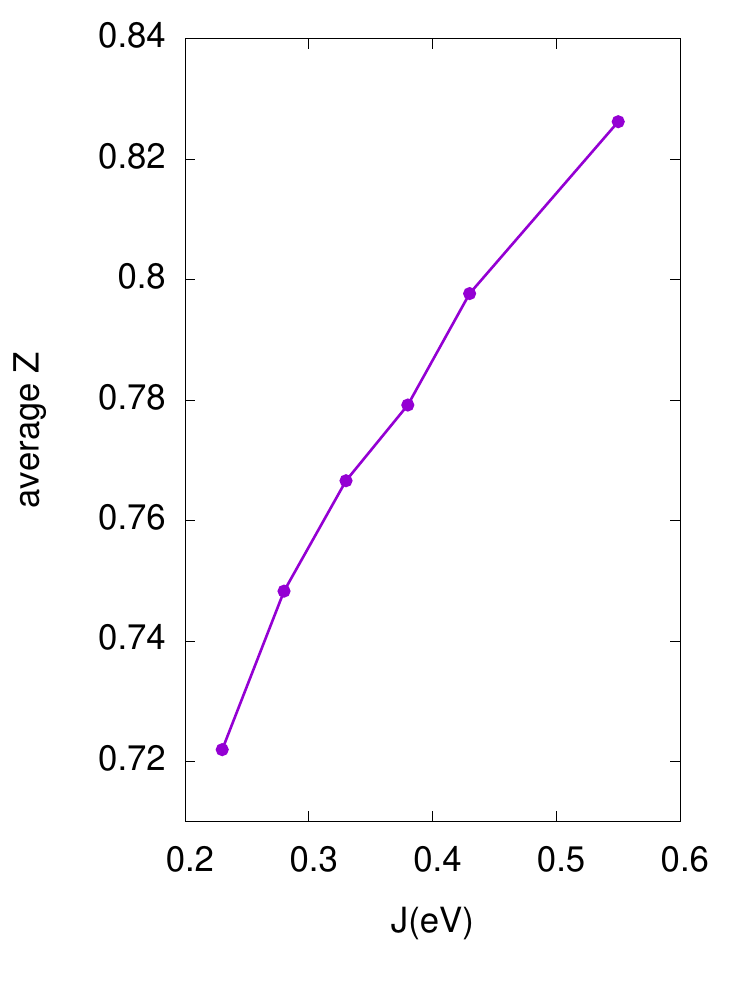} 
  \includegraphics[width=4.25cm]{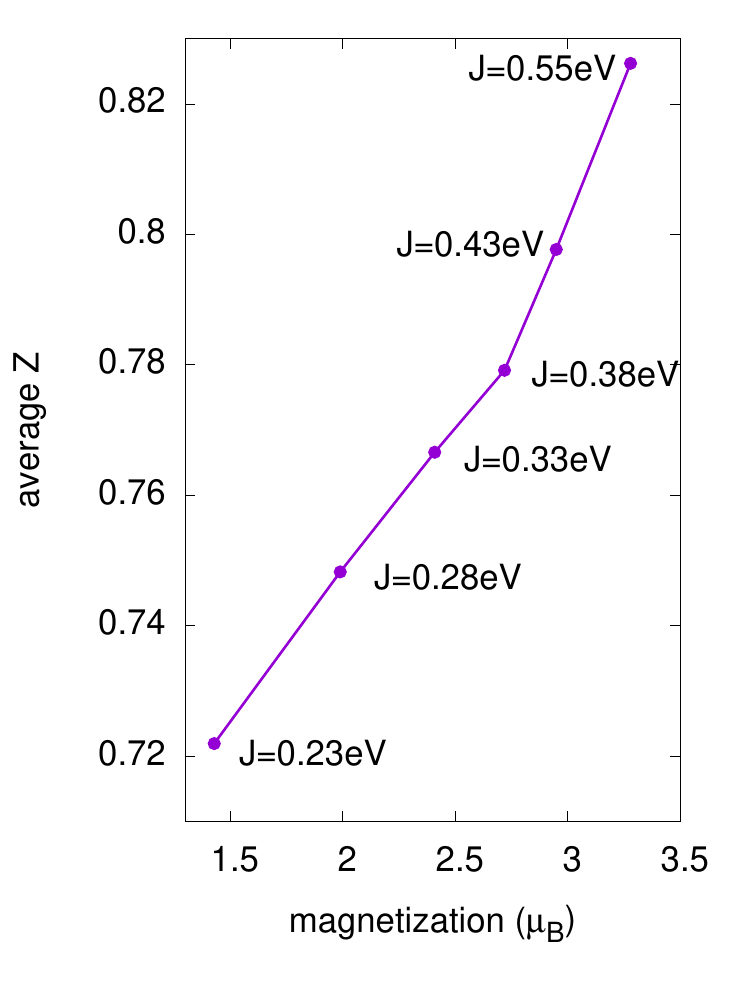} 
  \caption{Average (over spin and orbitals) quasiparticle weight $Z$ in the G-AFM phase within DFT+DMFT, for fixed $U=2.8$eV, as a function of $J$ (left panel) and of the corresponding magnetization (right panel). In this phase the system is decorrelated by Hund's coupling, as a consequence of a larger induced magnetization.}
  \label{fig:Zav_vs_J_mag}
\end{center}
\end{figure}

We have studied the dependence of the correlation strength on the value of Hund's coupling for fixed $U=2.8$eV, in the case of density-density interactions. The quasiparticle weight, averaged over spin and orbitals, is reported in Fig. \ref{fig:Zav_vs_J_mag}. Indeed one can see that the degree of correlation diminishes uniformly with increasing $J$ in this phase. The magnetization is also monotonically increased by a growing $J$.
Indeed when plotting the average $Z$ as a function of the magnetization one finds a rather linear behavior.
We can thus in general associate a larger magnetization to a reduced degree of mass renormalization. Even though we are in a metallic situation, this general result can be intuitively understood in the simple limit of a single-band fully spin polarized antiferromagnetic insulator, where the mass renormalization would be absent even if $U$ is extremely large.

The magnetization obtained for $U=2.8$eV and $J=0.43$eV with the Kanamori Hamiltonian is smaller (2.42$\mu_B$) compared to the density-density case (2.94$\mu_B$) and this supports the view according to which more pronounced magnetic effects reduce the quasiparticle effective mass renormalizaion.

\begin{figure}[h!]
\begin{center} 
  \includegraphics[width=8.5cm]{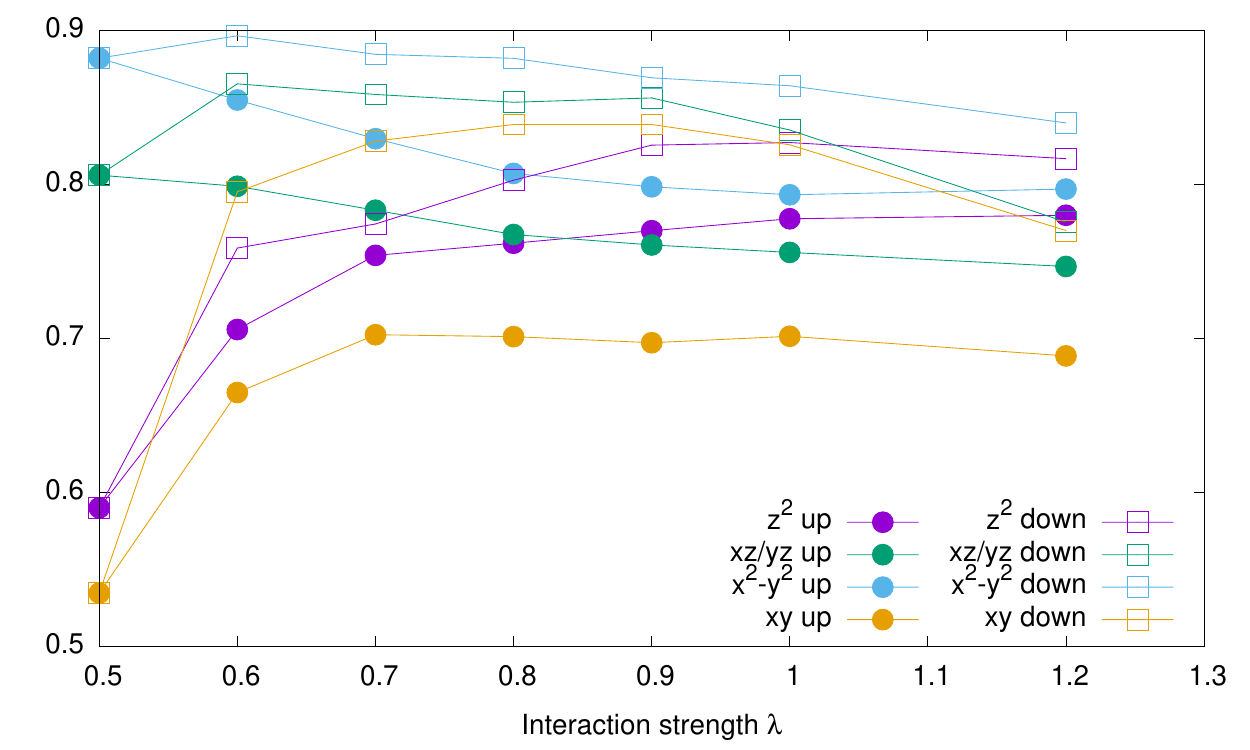} 
  \caption{Orbitally- and spin- resolved quasiparticle weights as a function of the interaction strength computed with DFT+DMFT at fixed $J/U=0.43$(eV)$/2.8$(eV)$=0.153$ ($\lambda=1$ for $U=2.8$eV and $J=0.43$eV) and $\beta=100$eV$^{-1}$.}
  \label{fig:Z_vs_lambda}
\end{center}
\end{figure}

\begin{figure}[h!]
\begin{center} 
  \includegraphics[width=8.5cm]{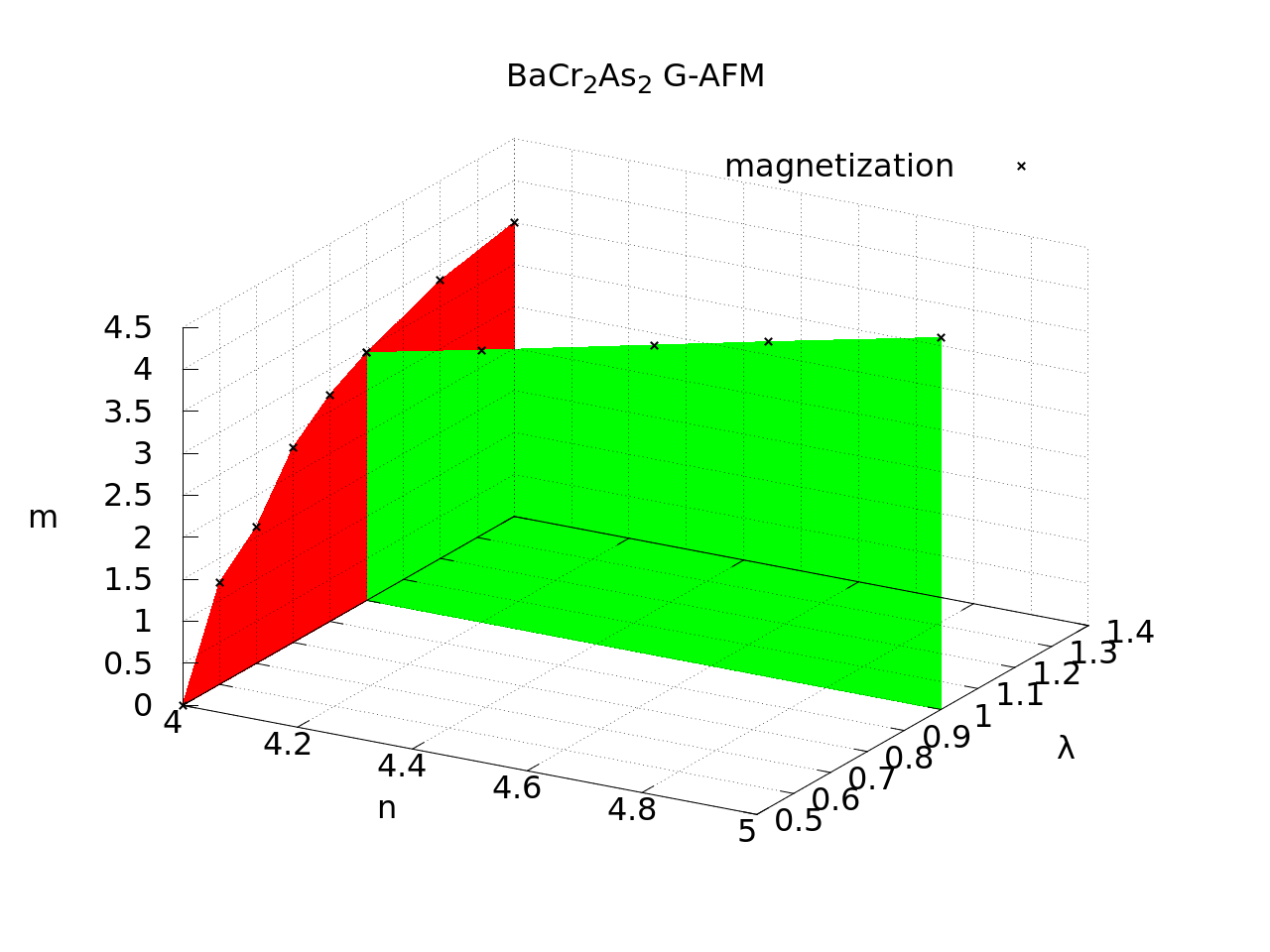} 
  \caption{Magnetization (in $\mu_B$) calculated with DFT+DMFT at  $\beta=100$eV$^{-1}$ for the density-density Hamiltonian in the plane density ($n$) - interaction strength $\lambda$. The large half-dome is symmetrically placed around half-filling ($n=5$), with respect to the analogous half-dome found between $n=5$ and $n=6$ found in Ref. \onlinecite{Misawa_d5-proximity_magnetic}.}
  \label{fig:magnetization_PhaseDiag}
\end{center}
\end{figure}

Having shed light on the quantitative difference between the results obtained with the two Hamiltonian, in the following we rely on the density-density form to perform a thorough exploration of the phase diagram of the magnetic phase. 
In Fig. \ref{fig:magnetization_PhaseDiag} we report the magnetization in the plane density-interaction (at fixed $J/U=0.43$(eV)$/2.8$(eV)$=0.153$). We parameterize, in the spirit of Ref. \onlinecite{Misawa_d5-proximity_magnetic}, the strength of the interaction through a parameter $\lambda$ which multiplies the full  $H_{int}$ so that $\lambda=1$ corresponds to the original values of the interaction $U=2.8$eV and $J=0.43$eV and $\lambda=0$ to a non-interacting system. The magnetization decreases by decreasing the interaction strength until it vanishes for $\lambda\simeq 0.5$. 
Interestingly the quasiparticle weights do not trivially rise towards $Z=1$, the non-interacting value, but show a plateau, as illustrated by Fig. \ref{fig:Z_vs_lambda}.
This is the outcome of the compensation between the decreasing interaction strength and the decreasing magnetization.

The magnetization is instead found to increase when increasing the electronic density away from $n=4$ and towards half-filling. This half-dome bears strong similarities to the one found in Ref. \onlinecite{Misawa_d5-proximity_magnetic} above half-filling for the iron pnictides. Indeed, besides decreasing with decreasing interaction strength, both stem from half-filling and decrease with doping away from it in a symmetric fashion.
Our (dynamical) mean-field treatment obviously overestimates the extension of the magnetic phase compared to the variational Monte Carlo treatment of Ref. \onlinecite{Misawa_d5-proximity_magnetic}, and so does the simplified density-density Hamiltonian, as mentioned.
Modulo these caveats the large extent in doping of this G-type antiferromagnetic dome stemming from the half-filled compound (which is a Mott insulator, see below) is a commonality between the present $d^4$ case and the $d^6$ physics of the pnictides that hints at the realization of similar many-body physics in the two cases.

\section{Paramagnetic phase: Hund's metal and compressibility enhancement} \label{paraSlSp}

In order to investigate more directly the many body physics realized in BaCr$_2$As$_2$ we turn now to its paramagnetic phase, where there is no magnetic ordering to compete with the effective mass renormalization and reduce the degree of correlation. 
As it is clear from the Z values for the lowest reported interaction value in Fig. \ref{fig:Z_vs_lambda} (that corresponds to the paramagnetic phase just outside the magnetic dome - see Fig. \ref{fig:magnetization_PhaseDiag}), this phase is therefore expected to be substantially more correlated than the magnetic one. 

\begin{figure}[h!]
\begin{center} 
  \includegraphics[width=4.25cm]{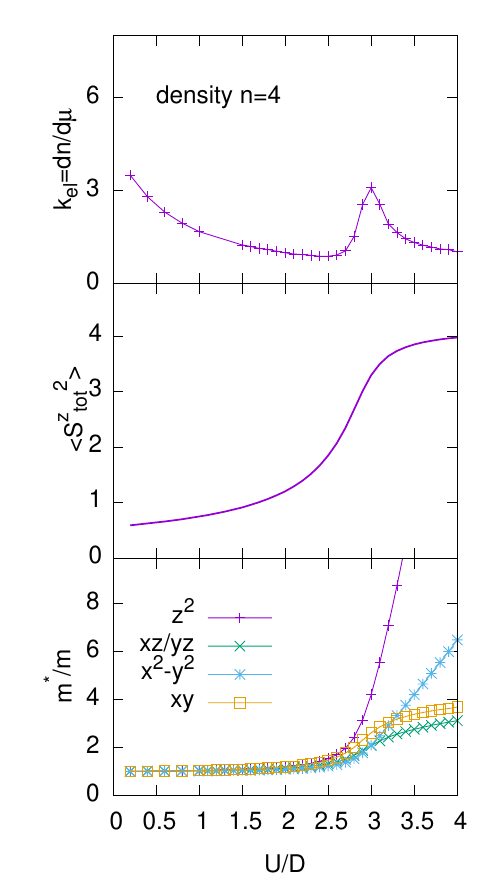} 
  \includegraphics[width=4.25cm]{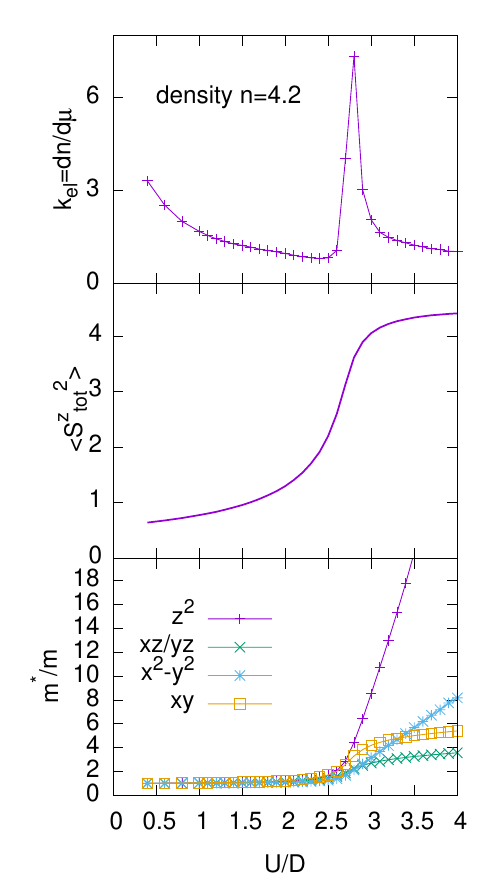} 
  \caption{Compressibility enhancement (top panels), total local magnetic moment and orbital-resolved mass enhancements as a function of $U$ ($J/U$=0.25) for constant densities $n$=4.0 (left) and $n$=4.2 (right). The compressibility enhancement is realized in correspondence of the crossover into the Hund's metal phase.}
  \label{fig:cross-over-peak}
\end{center}
\end{figure}

In order to fully explore the interaction-density phase diagram we use here the very agile and computationally cheap slave-spin mean-field method\cite{demedici_Vietri} (SSMF), which can be seen as a essential dynamical mean field capturing mainly the renormalization of the quasiparticle dynamics in the Fermi-liquid paramagnetic regime.
In this semiquantitative framework a fair agreement with DMFT using the Kanamori hamiltonian is obtained with the density-density Hamiltonian with a larger Hund's coupling compared to the physical value\cite{demedici_Vietri}. For BaFe$_2$As$_2$  $J/U=0.25$ was used (corresponding to a physical $J/U\simeq 0.15$, by comparisons between DMFT and SSMF in models\cite{demedici_Vietri}). This turns out to be a very satisfactory choice since it is capable to capture, with a single value of $U$, the quasiparticle renormalization\cite{demedici_OSM_FeSC} and specific heat Sommerfeld coefficient\cite{Hardy_122_SlaveSpin_exp} for the whole family of 122 FeSC.
We keep the same value in the present case, which also allows to compare the present chromium compounds with the iron counterparts.
\begin{figure}[h!]
\begin{center} 
  \includegraphics[width=8.5cm]{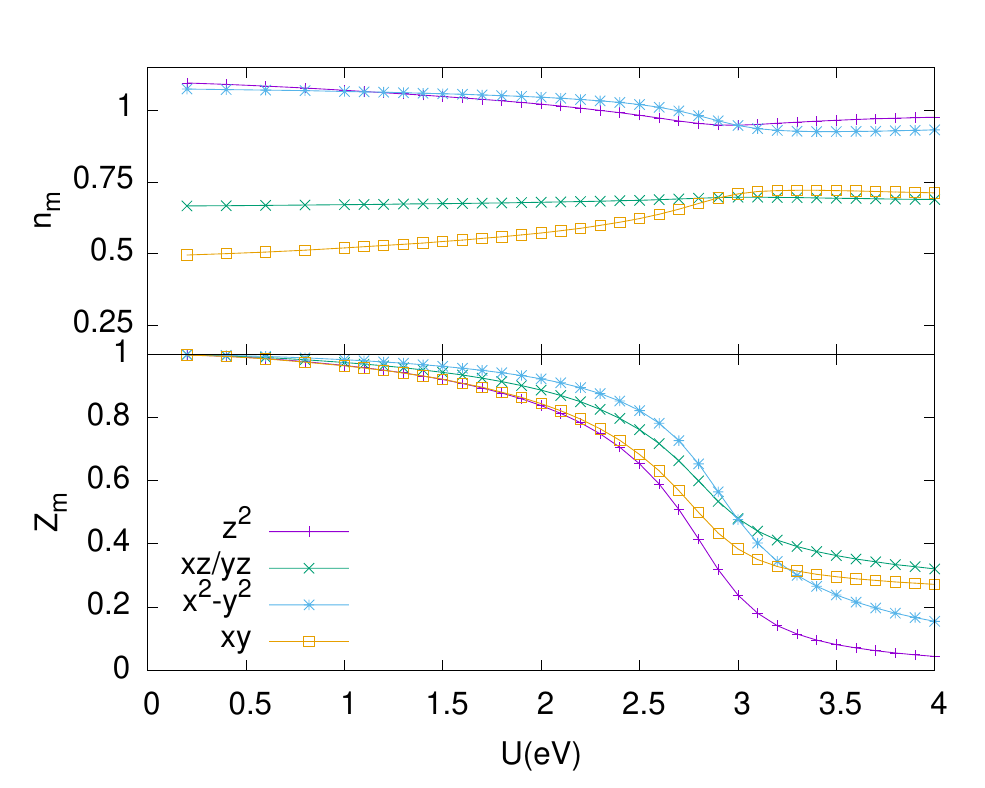} 
  \caption{Orbital populations and quasiparticle weight as a function of the interaction strength ($J/U=0.25$ is used as appropriate for BaFe$_2$As$_2$\cite{demedici_OSM_FeSC,Hardy_122_SlaveSpin_exp}) in the paramagnetic phase within slave-spin mean field.}
  \label{fig:Z_n_vs_U_cfr_DMFT}
\end{center}
\end{figure}
These calculations remarkably confirm our expectations that the many-body physics of this $d^4$ material is completely analogous to the $d^6$ case of BaFe$_2$As$_2$. Indeed one can observe the typical crossover\cite{Ishida_Mott_d5_nFL_Fe-SC,YuSi_LDA-SlaveSpins_LaFeAsO,Lanata_FeSe_LDA+Gutz,Yu_Si_KFeSe,Fanfarillo_Hund} between a low-$U$ weakly correlated phase where the quasiparticle weight (and correspondingly the mass enhancements, see lowermost panels in Fig. \ref{fig:cross-over-peak}) is very similar among all electrons, and a ``Hund's metal'' phase, which is much more strongly correlated and where the correlation strength is orbitally-selective.

In this respect it is to be noted that the $e_g$ orbitals ($z^2$ and $x^2-y^2$)  are more strongly correlated than the $t_{2g}$ orbitals ($xz$, $yz$ and $xy$) in this large-$U$ regime, opposite to what happens in FeSC. This is actually understood in term of the ``orbital-decoupling'' mechanism outlined in Refs. \onlinecite{demedici_MottHund,demedici_OSM_FeSC,demedici-SpringerBook,demedici_Vietri}.
This mechanism is based on the central role of the half-filled Mott insulator, which has a much smaller critical $U$ than any other integer filling, in the present case we find $U_c\simeq$2.4 eV. The proximity to this state 
decouples the charge excitations in the different orbitals, making each orbital behave like a single-band Mott insulator as far as the correlation dependence on the density is concerned: indeed, for each orbital $m$, $Z_m$ becomes linearly proportional to the respective orbital population $n_m$, with relatively similar slopes among the orbitals. As a main outcome the electrons in the different orbitals are more correlated the closer they are to individual half-filling. 

This is what explains the different values of $Z_m$ also in our theoretical study of BaCr$_2$As$_2$. The lower overall electron density with respect to BaFe$_2$As$_2$ moves the $e_g$ orbitals closer to individual half-filling than the $t_{2g}$ (upper panel in Fig. \ref{fig:Z_n_vs_U_cfr_DMFT}), opposite to what happens in BaFe$_2$As$_2$. This naturally explains the inverted correlation stregnth between the two kind of orbitals in the compounds at large $U$.

As outlined previously for FeSC\cite{Lanata_FeSe_LDA+Gutz,Hansmann_localmoment_prl,demedici_Vietri,demedici_Compressibility} this crossover is characterized also by a sharp growth of the instantaneous local moment (see middle panels in Fig. \ref{fig:cross-over-peak}), which reflect that the strong-coupling metallic phase is dominated by the high-spin configurations favored by Hund's coupling, motivating the name ``Hund's metal''.

Another remarkable feature developing at the crossover was recently reported in Ref. \onlinecite{demedici_Compressibility}: a strong enhancement of the electronic compressibility $\kappa_{el}=\frac{dn}{d\mu}$  just inside the frontier of the Hund's metal phase. In a density-interaction plane, the region where the enhancement takes place departs from the half-filled Mott transition like the Hund's metal crossover, and extends at finite doping reaching larger values of $U$.
At low doping, where the crossover is sharper, this compressibility enhancement culminates with an actual divergence, while it slowly fades away at large doping. 
 
 This finding is again confirmed here for BaCr$_2$As$_2$, as shown in the upper panel of Fig. \ref{fig:cross-over-peak}, where the correspondence with the Hund's metal crossover is highlighted.
 
\begin{figure}[h!]
\begin{center} 
  \includegraphics[width=8.5cm]{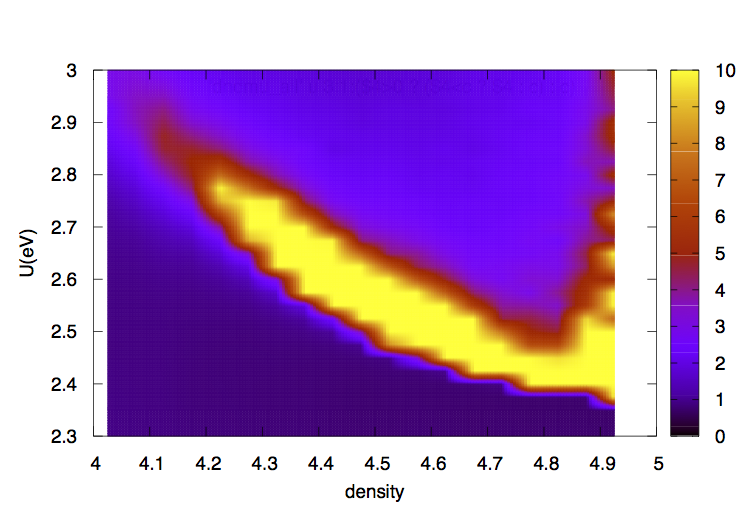} 
  \caption{Compressibility calculated for BaFe$_2$As$_2$ in the interaction-density plane within SSMF ($J/U=0.25$). Solid yellow color indicates values of $\geq$ 10 and the actual instability (divergent or negative values).The actual compound is located at density $n$=4 and $U$=2.8eV or slightly less, so that negative pressure and/or electron doping should bring it inside the enhancement zone, possibly favoring hight-temperature superconductivity.}
  \label{fig:Compressibility}
\end{center}
\end{figure}

In Fig. \ref{fig:Compressibility} a full map $\kappa_{el}$ is reported in the interaction-density plane.
This shows that the zone of compressibility enhancement extends until the density $n$=4 relevant for BaCr$_2$As$_2$, symmetrically to the case of BaFe$_2$As$_2$ where it extends until $n$=6\cite{demedici_Compressibility}.
Differently from that case however, we find here that the estimated interaction strength for the compound ($U=2.8$eV or slightly less), places it below the zone of enhancement. We thus speculate that chemical substitution with smaller ions (negative pressure) reducing the ratio between $U$ and the hopping matrix elements,  or more directly electron doping, might bring the compound within the zone of  enhancement.

The interest of this compressibility enhancement lies in the fact that it is the outcome of an anomaly in the residual quasiparticle interactions\cite{demedici_Compressibility}. 
This might favor superconductivity by directly driving an interaction between quasiparticles (testified by a negative Landau parameter $F_0^s$) as well as a booster of some electron-boson vertices involved in the pairing mechanism.

Provided thus the aforementioned chemical substitutions are sufficient to get rid of the magnetic ordering of the parent compounds, they might actually favor high-temperature superconductivity.

\section{Conclusions}

In this work we have theoretically explored both the experimentally realized\cite{Singh_BaCr2As2} G-type magnetic metallic phase of BaCr$_2$As$_2$, and the corresponding metallic phase where this order has been suppressed.
In the magnetic phase DFT+U and DFT+DMFT calculations allowed us to show that dynamical correlations are necessary to obtain a specific-heat Sommerfeld coefficient close to experiments. In this framework we have clarified that the formation of an ordered magnetic moment reduces the degree of correlation and leads to a relatively small effective mass renormalization factor $\sim$2 despite the estimated substantial value of the interaction. We thus find that at the same interaction strength the paramagnetic phase is more correlated.

In this phase, using SSMF to thoroughly explore the whole phase diagram in the interaction-density plane using the electronic structure of BaCr$_2$As$_2$, we establish that the many-body physics in this compounds mirrors that of BaFe$_2$As$_2$, motivating also the corresponding similarity of the magnetic phase diagrams of the two materials.

Indeed the phase diagram features a crossover from a weakly correlated conventional metal to a strongly correlated metal with orbital-selective quasiparticle weights is found as a function of $U$, on a frontier departing from the half-filled Mott transition and bending towards larger $U$ at finite doping. Analogously to the FeSC case we also find a compressibility enhancement accompanying this frontier.

Assuming for BaCr$_2$As$_2$ a similar interaction strength as for BaFe$_2$As$_2$, the compound  lies close and slightly below the region of enhancement. We can conjecture that negative pressure and/or electron doping might bring it right in the zone of compressibility enhancement, favoring the appearance of electronic instabilities, including superconductivity. This could be realized by e.g. substituting chromium with manganese or iron  or replacing barium with lanthanum, scandium or yttrium.

We remark that the physics that we have outlined in this paper stems largely from the $d^4$ filling of the five $d$-orbitals of chromium, and doping away from it. This implies that our conclusions might be relevant for other compounds of chromium (such as for instance K$_2$Cr$_3$As$_3$\cite{Bao_K2Cr3As3}) and in general for similar compounds with filling in the proximity of $d^4$, if a Mott insulator is realized at half-filling in the relevant band structures.
 
\acknowledgements We thank G. Giovannetti for his help with the Wannier projection in the early stages of this work. Useful discussions with A. Hausoel, M. Karolak, L. Fanfarillo and A. Toschi are also acknowledged. 
The authors gratefully acknowledge the Gauss Centre for Supercomputing e.V. (www.gauss-centre.eu) for funding this project by providing computing time on the GCS Supercomputer SuperMUC at Leibniz Supercomputing Centre (LRZ, www.lrz.de).

%

\end{document}